\documentclass{article}
\usepackage{epsfig}
\usepackage{multirow}
\usepackage{amsmath}

\usepackage{amssymb}
\def\f{\frac}
\def\z{\noindent} 
\def\p{\partial}
\def\app{\approx}
\def\l{\lambda}

\begin{document}
\date{}

\title{Minimum current principle and variational method in theory of space charge limited flow}

\author{A.Rokhlenko 
\medskip
\\Department of Mathematics,
Rutgers University\\ Piscataway, NJ 08854-8019}

\maketitle
\begin{abstract}

In spirit of the principle of least action, which means that when a perturbation 
is applied to a physical system its reaction is such that it modifies its state to 
"agree" with the perturbation by "minimal" change of its initial state. In particular, 
the electron field emission should produce the minimum current consistent with 
boundary conditions. It can be found theoretically by solving corresponding 
equations using different techniques. We apply here for the current calculation 
the variational method, which can be quite effective even when involving a short set
of trial functions. The approach to a better result can be monitored by the total
current that should decrease when we on the right track. Here we present only an
illustration for simple geometries of devices with the electron flow. The development
of this methods can be useful when the emitter and/or anode shapes make difficult 
the use of standard approaches. Though direct numerical calculations including PIC 
technique are very effective, but theoretical calculations can provide an important 
insight for understanding general features of flow formation and even sometimes be 
realized by simpler routines. 

\end{abstract}

\z PACS: 52.59.Sa; 79.40.+z; 85.45.Bz; 02.30.Jr 

\bigskip
We implement here the variational method whose efficiency verified by the current 
minimization for a few simple cases of the electron flow produced by the field emission 
in model diodes when the initial speed of electrons is negligible. The electric potential 
in our diodes is $\varphi(\vec r)$, space charge density $\rho(\vec r)$, current density 
$\vec j(\vec r)$, and the electron velocity $\vec v(\vec r)$. The electron mass and the
absolute value of its charge are $m$ and $e$. Assuming a well conducting emitter the
electric field $\vec f(s)$ is normal to each point of its surface. In this work we
construct the variational methods of finding the currents, specific for each considered
set up and compare them with the exact ones or reliable approximations when the later 
are known. The approach to better approximations is verified by using the principle of
minimum stationary current as a special case of the universal principle of least action 
discussed in particular by R. Feynman in [1] and other works.

\medskip\noindent
{\bf 1. Child-Langmuir Flow}
\medskip

The Child-Langmuir (CL) current [2], [3] is the simplest case of a one-dimensional (1D) 
space charge limited electron flow in a diode, especially in the planar system with cathode 
at $x=0$ and anode at $x=d$ which can be easily treated analytically by solving the Poisson 
equation$$
\f{d^2\varphi(x)}{d x^2}=\f{\rho(x)}{\epsilon_0}=\f{j}{\epsilon_0 v(x)}=\f{j}{\epsilon_0}
\sqrt{\f{m}{2e\varphi(x)}}\eqno(1)$$
with the boundary conditions (BC)$$
\varphi(0)=0,\ \ \varphi(d)=V,\eqno(2)$$
where $\epsilon_0$ is the vacuum permittivity.
A simple observation shows that Eq.(1) can be viewed as the Euler equation for 
the variational functional [4], [5]$$
M=\int_0^d\left\{\f{1}{2}[\varphi'(x)]^2+2\f{j}{\epsilon_0}\sqrt{\f{m\varphi(x)}{2e}}
\right\}dx.\eqno(3)$$

Let us show that in the CL systems the solution of Eqs.(1,2), $\varphi(x)=V(x/d)^{4/3}$, is 
the minimizer of $M$. We begin by taking the simple class of trial functions $u(x)=
V(x/d)^\l$ where $1<\l<2$ because $\varphi''(x)\propto\rho(x)$ in Eq.(1) is infinite at 
$x=0$, but the field $\varphi'(0)=0$. 

First we substitute $u(x)$ with its BC $u(0)=0,\ u(d)=V$ in Eq.(3) instead $\varphi(x)$ 
$$M=\f{V^2\l^2}{d(2\l-1)}+\f{jd}{\epsilon_0}\sqrt{\f{mV}{2e}}\f{4}{\l+2},\eqno(4)$$
compute the $\l$-derivative of $M$ assuming $j$ fixed$$
\f{dM}{d\l}=\f{V^2 l(\l-1)}{d(2\l-1)^2}-\f{jd}{\epsilon_0}\sqrt{\f{mV}{2e}}\f{4}
{(2+\l)^2},\eqno(5)$$
then using the current density $j=\rho(x)v(x)$ from (1) find its average in the diode$$
\bar{j}=\f{\epsilon_0}{d}\int_0^d{\f{d^2u}{dx^2}(x)\sqrt{\f{2eu(x)}{m}}dx}=\epsilon_0V^{3/2}
\sqrt{\f{2e}{m}}\f{2\l(\l-1)}{d^2(3\l-2)}.\eqno(6)$$
Evaluating the derivative (5) after substitution Eq.(6) for $j$ one finds$$
\f{dM}{d\l}=V^2d^{-1}\l(\l-1)\left[\f{1}{(2\l-1)^2}-\f{8}{(\l+2)^2(3\l-2)}
\right]\eqno(7)$$ 
which becomes zero when $\l=4/3$ ($\l=0,\ 1$ contradict the BC). The second derivative at 
this point$$
\f{d^2M}{d\l^2}=V^2d^{-1}\left[\f{1}{2\l-1)^3}+\f{16\l(\l-1)}{3\l-2)(\l+2)^3}\right]$$
is positive and therefore we have for $\l=4/3$ the minimum of functional $M$. The diode
current density (6) with this $\l$ is the well known CL result$$
j_{CL}=\f{4}{9}\sqrt{\f{2e}{m}}\f{\epsilon_0 V^{3/2}}{d^2}.\eqno(8)$$
The minimum is narrow: when $\l=4/3\pm 10\%$ the factor $4/9$ changes by $\sim 30\%$.
This phenomenon might indicate that, while the general potential shape in a system can be
computed quite well using the minimization of $M$, the current is more sensitive to 
approximations. To check up this we studied a simple two-term model of the potential $u(x)$
with incorrect behavior at $x=0$
$$\sqrt{u(x)}=a\left(\f{x}{d}\right)^{3/4}+(\sqrt{V}-a)\f{x}{d},\eqno(9)$$
where $a$ should be found. The minimization procedure, similar to one above, yields $a=
1.17916\sqrt{V}$ and the average on $(0,d)$ current density,$$
\bar{j}=0.5175\sqrt{\f{2e}{m}}\f{\epsilon_0 V^{3/2}}{d^2},\eqno(10)$$
is $\sim 16\%$ higher than the exact $j$ in Eq.(8). In approximations the current density 
$j$ clearly cannot be constant as a function of $x$, it is even equals to zero at $x=0$. 
This affects the precision of calculation and the direct minimization of ${\bar j}$ with 
this behavior of $j(x)$ is impossible. The average difference between $\varphi(x)$ and 
this $u(x)$ on the interval $(0,d)$ is calculated to be only $1.7\%$. 

Note that this analysis represents a suggestive illustration for validity of our claim. A 
very simple way of solving this problem comes from the requirement of the current density
independence of $x$ in the stationary state: $j=\varphi''(x)\sqrt{\varphi(x)}=const$, i.e. 
immediately $\l=4/3$. Our approach assumes wider applications of the minimum principle,
especially when one needs the current evaluation in higher dimensions.

\medskip\noindent
{\bf 2. Electron flow in 1D diode when cathode field $f\neq 0$}
\medskip

The potential in this system is governed by Eq.(1), but $\varphi'(0)=f>0$ instead of BC (2). 
Eq.(1) nevertheless can be solved analytically [6] in the stationary state and this solution, 
which involves $\varphi(x)$ and the current density $j$, is described by the following equation$$
j(x)=\f{\epsilon_0}{9x^2}\sqrt{\f{2e}{m}}\left\{2\varphi^{3/2}(x)+[2\varphi(x)-3xf]
\sqrt{\varphi(x)+3xf}\right\},$$
when $v(0)=0$. In particular, if the anode voltage is $V$ and cathode-anode distance
is $d$, the relationship between the current density and the cathode electric field $f\geq0$, 
derived in [2], in physical units has the following form$$
j=\f{\epsilon_0}{9d^2}\sqrt{\f{2e}{m}}\left[2V^{3/2}+(2V-3df)\sqrt{V+3df}\right].\eqno(11)$$
Eq.(11) is universal independent of the emission law $j=j(f)$. When $j(f)$ is known 
it can be substituted into Eq.(11) which can be solved then for the current evaluation in this 
system. Note that in a free of space charge diode $fd=V$ and $j=0$ in agreement with Eq.(11). A 
simple analysis shows also that in Eq.(11) $j$ decreases faster than $d^{-2}$ (when $V$ and $f$ 
fixed) and increases faster than $V^{3/2}$ ($d$ and $f$ fixed) compared with $j_{CL}$. Clearly
$j(x)$ is the same stationary current density at any arbitrary cross-section of this diode, $dj/dx
=0$.

For simplification we proceed now in dimensionless variables with $V=d=1$, $m=2$, and $
\epsilon_0\sqrt{e}=1$. Eq.(11) takes the following form:$$
j=[2+(2-3f)\sqrt{1+3f}]/9.\eqno(12)$$
In particular, for $f=0.2,\ 0.4, 0.6,\ 0.8$ Eq.(12) yields $j=0.419,\ 0.354,\ 0.259,
\ 0.140$ respectively. 

In the approximate study of this problem we use a simple model for the potential $\varphi(x)=
fx+ax^{3/2}+(1-a-f)x^2$ with a single free parameter $a$. This form for  $\varphi(x)$ reflects 
a finite electric field $\varphi'(0)=f$ at the cathode surface and an infinite value of 
$\varphi''(0)$ necessary for a non-zero current density $j=\varphi''\sqrt{\varphi}$ at $x=0$ when 
the initial velocity $\sqrt{\varphi(0)}$ assumed negligible. Then we construct the functional $M$ 
using Eq.(3) and minimize it by finding for each $f$ such an $a$ which makes the first derivative 
of $M$ zero ($j(a)$ is not differentiated). After finding $a$ and substituting it into $j(a)$ we 
compare $\bar j$ with the exact values of $j$ from Eq.(12). For the four quantities of $f$ above 
the averages current densities
$$\bar{j}=\int_0^1{j(x)dx}$$
are $0.484,\ 0.387,\ 0.273,$ and $0.143$ respectively, witch are higher (as they should) than the 
corresponding exact values by only $15.4,\ 9.4,\ 5.3,\ 2.3$ percents. The precision is better 
than in the CL case because in modeling the current we take care about $j(x)$ behavior inside 
the diode. Note that one cannot look directly for the minimum of $j$ with this crude model of 
$\varphi(x)$ which even makes $\varphi''=\rho<0$ when $a>1.6(1-f)$.

Results presented in parts 1 and 2 show that approximate current densities in 1D problems are
always higher than the exact ones. This confirms the principle of minimum current in stationary
systems.

\medskip\noindent
{\bf 3. Treating flow in higher dimensions}
\medskip

The generalization [4] of the variational functional in three dimensions has the following form$$
M=\int_{\Omega}\left\{\f{1}{2}\left [\left( \f{\p\varphi}{\p x}\right)^2+\left( \f{\p\varphi}{\p y}
\right)^2+\left( \f{\p\varphi}{\p z}\right)^2\right ]+2j(x,y,z)\sqrt{\varphi(x,y,z)}\right\}dxdydz,
\eqno(13)$$
where $\Omega$ is the volume occupied by the current $j$. The potential $\varphi(x,y,z)$ should
satisfy the BC on the volume boundaries. The minimum of $M$ is realized by the solution of the
Euler-Lagrange-Ostrogradsky equation which generalizes Eq.(1),$$
\triangle\varphi(x,y,z)=\rho(x,y,z)=\f{j(x,y,z)}{v(x,y,z)},\eqno(14)$$
in the case when the current density $j(x,y,z)$ is a given function. It represents the standard 
Poisson equation for the potential in the region $\Omega$. Eq.(14) together with its BC in principle 
allow to find the local current density and electron velocity in terms of $\varphi$ using the current 
continuity relation, but Eq.(13) agrees with Eq.(14) only when $j(x,y,z)$ is fixed and therefore one
cannot use the minimizer $\varphi$ of $M$ straightforwardly as the shape of $j(x,y,z)$ is determined 
by $\varphi$. In the one-dimensional case $j$ is an unknown constant unused in the minimization 
routine, but after it $j$, expressed in terms of parameters of $\varphi(x)$, is substituted 
in the derivatives of $M$ and then calculated, see Parts 1 and 2. 

This represents a difficulty in higher dimensions, where a model for $j(x,y,z)$ should be substituted
in Eq.(13). On the other hand, it is clear that with given electrode potentials an incorrect current 
density might violate the BC (like emissivity) at the cathode, as it was in the 1D case when we 
used the trial function (9).

For illustration we consider here the two-dimensional system, where the cathode and anode are 
parallel planes but the cathode has smooth periodic bumps analogous to [7]. The calculation can 
be done within a single cell whose width in dimensionless units is $2$ and distance between the 
valley of the cathode and anode is $d$, see Fig.1. 
\vskip0.2cm
\hskip3cm
\epsfig{file=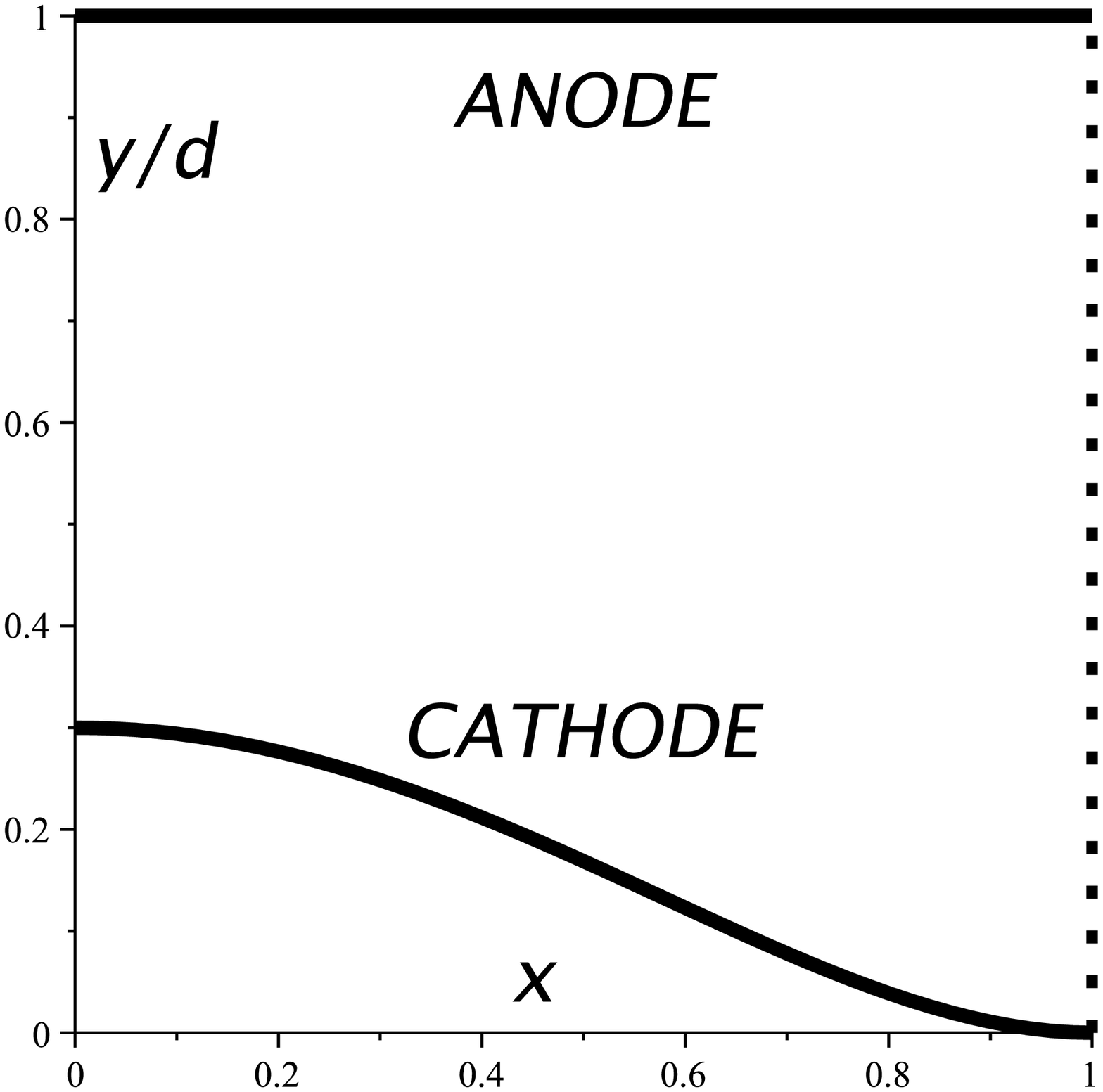, width=4cm, height=4cm} 

\centerline{\small Fig.1. Cathode bump represented by $u(x)$ in a half-cell of 2-D diode} 
\bigskip\noindent
The shape of the bump is described by the function$$  
u(x)=a(1-x^2)^2,\ \ a\geq 0.\eqno(15)$$
A strong magnetic field in the $y-$direction makes the electron trajectories vertical and therefore 
the current density depends on $x$ only, $j=j(x)$. In the case of the infinite cathode emissivity,
anode voltage $V$, and $a=0$ the CL law gives$$
j_{CL}=\f{4V^{3/2}}{9d^2}.\eqno(16)$$ 
In [7] for approximating $\varphi(x,y)$ we used about 20-25 trial functions and obtained for $a=0.3$
the average over $x$ current density $\bar j\app 1.596j_{CL}$. 

Here we model  $\varphi$ by the following sum$$
\varphi(x,y)=\sum_{i=1}^N{c_i\left[\f{y-u(x)}{d-u(x)}\right]^\f{2i+2}{3}},\ {\rm where} 
\ \f{\p \varphi}{\p x}(0)=\f{\p \varphi}{\p x}(1)=0.\eqno(17)$$ 
Such a form can create difficulties in integrating $\sqrt{\varphi}$ when the coefficients $c_i$ are
unknown, but for $N>1$ we use iterations which means that the coefficients in  $\sqrt{\varphi}$ are 
taken from the previous step ($c_i\to b_i$) and the integral in a higher dimension of Eq.(13) is just 
a number as all $b_i$ are known. It is important to comment that in Eq.(13) $b_i,\ i=1..N$ enter via 
the integral of $\sqrt{\varphi}$ in (13) which is not very sensitive to these coefficients, 
therefore after $c_i$ are found we can replace $b_i=c_i,\ i=1..N$ and go to the next step of 
iterations.

The CL law (16) in the particular case of Eq.(17) can be realized in the differential form for 
the cathode of infinite emissivity when $y$ approaches to $u(x)$. The $y-$component of the
current density in the form of Eq.(16) in view of (17) is determined by the first term $i=1$
only:
$$j_y(x)=\f{4[\varphi(x,y)]^{3/2}}{9(\Delta y)^2}=\f{4}{9[y-u(x)]^2}\left [c_1\left(\f{y-u(x)}
{d-u(x)}\right)^{4/3}\right]^{3/2}=\f{4c_1^{3/2}}{9[d-u(x)]^2}.$$
The strong magnetic field makes the following form of the total current$$ 
j(x)=\f{4c_1^{3/2}\sqrt{1+[u'(x)]^2}}{9[d-u(x)]^2},\eqno(18)$$ 
where the square root in Eq.(18) converts the $y-$derivative into the normal derivative to a 
surface element. Clearly when $\varphi(x,y)$ is modeled differently one would obtains a 
different equation, but the idea of finding a proper form corresponding to Eq.(18) is obvious.

After integrating $\left( \f{\p\varphi}{\p x}\right)^2+\left( \f{\p\varphi}{\p y}\right)^2$ and 
finding the partial derivatives in $c_1...c_N$ we add to each of them the corresponding integrals$$
\int_0^d{j(x)dx}\int_{u(x)}^d{\f{\p \sqrt{\varphi}}{\p b_k}dy},\ \ k=1..N,\eqno(19)$$
calculated using Eqs.(13) and (18). The minimization procedure means that these sums of derivatives 
are zeros. This allows to evaluate $c_i,\ i=1..N$ and go to the next step. The process is quite 
simple and the convergence up to $3-4$ digits needs only a few iterations.

In the present computations taking $a=0.3$ and measuring the current density in the CL units as 
$j_{tot}=\bar j/j_{CL}$ we get for $N=1,2,3$ by Eq.(17) the following total currents $j_{tot}=
1.559,\ 1.449,\ 1.444$ respectively. They differ from the result of [7] by $\leq 10\%$ and 
decreasing with the improvement of modeling $\varphi(x,y)$. This supports our claim about the 
current minimization. The addition of the third term in Eq.(17) is clearly ineffective. Note 
that there are only $N-1$ independent coefficients $c_i$ because the BC at the anode requires 
$\sum{c_i}=V$. Applying the present method we get decent results in calculating the current 
density with only $2-3$ trial functions and using a simpler technique. 

Treating the magnetized flow in three dimensions like in [8], when the emitter has periodic 
pattern of bumps can be performed by this method in a similar way after choosing a proper set
of trial functions.

\medskip\noindent
{\bf 4. Conclusion}
\medskip

1. A principle of minimum emitted current in stationary systems is offered and illustrated in a 
few examples. Its wider and effective applications wait for attention of researchers.

2. We use here the effective variational method, see Eqs.(3) and (13), for solving the field 
emission problems in planar and simple two-dimensional geometries. This method can be generalized.

3. The differential form of the Child-Langmuir law (Eq. (18) in our case) is derived an applied
for the two-dimensional problem studied in section 3. 

4. For treating the flow in 2 and 3 dimensions, when the integrals in Eq.(19) cannot be evaluated 
analytically, a very effective iteration procedure is implemented.

\bigskip\noindent
{\bf ACKNOWLEDGMENTS}
\bigskip 

The motivation of this study was influenced by our discussions with H. Brezis to whom I 
express my thanks. I also benefited from conversations and encouragement by J. L. Lebowitz.

\end{document}